\documentclass[12pt]{article}

\usepackage{amsmath}
\usepackage{latexsym}
\newcommand\be{\begin{equation}}
\newcommand\ee{\end{equation}}
\newcommand\bea{\begin{eqnarray}}
\newcommand\eea{\end{eqnarray}}
\newcommand\beas{\begin{eqnarray*}}
\newcommand\eeas{\end{eqnarray*}}

\def\tr{{\rm Tr}}

\begin{document}
\title{Large $N$ Spectrum of two Matrices in a Harmonic Potential and BMN energies}

\author{\\
Jo\~ao P. Rodrigues \\
\\
Department of Physics\\
Princeton University\\
Princeton, NJ 08544, USA\\
 \\
School of Physics and Centre for Theoretical Physics \\
University of the Witwatersrand\\
Wits 2050, South Africa \footnote{Permanent address. 
Email: rodriguesj@physics.wits.ac.za}\\
}

\maketitle

\begin{abstract}
\end{abstract}
The large $N$ spectrum of the quantum mechanical hamiltonian of two hermitean matrices 
in a harmonic potential is studied in a framework where one of the matrices is treated 
exactly and the other is treated as 
a creation operator impurity in the background of the first matrix. For the free case,
the complete set of invariant eigenstates and corresponding energies are obtained.
When $g_{YM}^2$ interactions are added, it is shown that the full string tension corrected 
spectrum of BMN loops is obtained.

\newpage

\noindent
\section{Introduction}

Based on studies of giant gravitons in AdS x S and their SYM duals [1-8], a dual description of 1/2 BPS states has been 
arrived at in terms of free fermions associated with a complex matrix in a harmonic potential \cite{Corley:2001zk},\cite{Berenstein:2004kk}. 
More recently, it was established in \cite{Lin:2004nb} that the gravity description of 1/2 BPS states is completely determined  
by a density function associated with a general fermionic droplet configuration. Many other authors have
studied this correspondence [11-34].

However, as emphasized recently in \cite{Donos:2005vm}, the energy and flux of the 1/2 BPS states obtained in \cite{Lin:2004nb}
are also those of a one dimensional hermitean matrix, in a bosonic phase space density description. 
In \cite{Donos:2005vm}, an extension of 1/2 BPS states was studied by first treating this hermitean matrix exactly, 
and then by studying fluctuactions of states with "impurities" of the other matrix (in a creation/annihilation basis) 
about the large N background of the first matrix.    
A discrete linear spectrum of states was obtained, and a map to gravity states with either S or AdS radial dependence was proposed. 

This work builds on earlier work \cite{deMelloKoch:2002nq},\cite{deMelloKoch:2003pv} 
where an approach to the derivation of string field theory from matrix model hamiltonians 
in the BMN limit was developed: in \cite{deMelloKoch:2002nq}, the hamiltonian of free matrices in a harmonic potential resulting 
from compactification of scalars on 
$S^3 \times R$  was considered, and the spectrum and cubic interactions of supergravity modes were obtained. 
The background was generated from known two point functions and the use of Schwinger-Dyson equations. 
In \cite{deMelloKoch:2003pv}, a matrix model Hamiltonian
suitable to the pp wave limit in a creation/annhiliation basis was proposed that, once operator mixing is taken into account, 
was shown to correctly reproduce the pp wave light cone string field theory. Some knowledge of the properties of the
underlying degrees of freedom is required.    

The framework used in  \cite{Donos:2005vm},\cite{deMelloKoch:2002nq},\cite{deMelloKoch:2003pv} is based on a change of variables 
from the original matrix degrees of 
freedom to gauge invariant collective matrix fields \cite{Jevicki:1979mb}, whose spectra and interactions are susceptible of a 
gravity/string theory interpretation. This approach was used successfully in the developement 
of a string field theory of c=1 strings \cite{DasKA},\cite{Demeterfi:1991tz}. 

In this article I investigate further the proposal of \cite{Donos:2005vm} that the large N properties of multi matrix models can
be studied by solving one of the matrices exactly in the large N limit and by treating the other matrices in the backgound 
induced by that matrix in a systematic way.

A first check of this approach is to ensure that the full spectrum of gauge invariant states of two free hermitean matrices in a 
harmonic potential is recovered. This is carried out in this article, by generalizing the states considered in \cite{Donos:2005vm}.

As pointed out in \cite{Donos:2005vm}, one can think of the "impurity" as associated with the antiholomophic complex matrix, but one
can also associate it with one of the scalars in the transverse directions to the complex matrix plane. I elaborate on this last aspect of the problem. 
By identifying the analogue BMN \cite{Berenstein:2002jq} states and by carrying perturbation theory in $g_{YM}$, I 
obtain the first order string tension corrected BMN spectrum \cite{Berenstein:2002jq}. 
After performing a Bogoliubov transformation,
I obtain the full string tension corrected BMN spectrum 
\cite{Santambrogio:2002sb},\cite{Berenstein:2005jq}.

This paper is organized as follows: in Section $2$, the large N limit of the quantum mechanics of two hermitean matrices in a harmonic potential is 
studied, with one of the matrices (the "second matrix") being treated in a creation/annihilation basis. The most general set of gauge invariant 
states is identified, and after treating the first matrix exactly, the leading and quadratic hamiltonian in terms of these gauge invariant
states is arrived at \cite{Jevicki:1979mb},\cite{Avan:1995sp},\cite{Donos:2005vm}. 
The equation determining the spectrum of states with impurities is found to be a multi-local generalization
of the Marchesini Onofri equation \cite{MarchesiniYQ},\cite{Halpern:1981fj},\cite{Gross:1990md},\cite{Boulatov:1991xz},\cite{Maldacena:2005hi},\cite{Donos:2005vm}, 
which appears in studies
of the non-singlet sector of the single matrix theory. The spectrum is linear, with a large degeneracy.
Chebyshev polynomials of the second kind
play a special role in the description of the eigenfunctions.     

In Section $3$, the result of introducing $g_{YM}$ interactions is discussed. This first requires an identification of the original matrix variables
used in Section $2$. The identification of 
matrix variables implied by the 1/2 BPS state description is reviewed, and I obtain the form of a typical $F$ term in these variables. 
Writing down the analogue BMN \cite{Berenstein:2002jq} loops by quantum number matching, perturbation theory is carried out and the 
the first order $g_{YM}$ corrected oscillator spectrum is obtained.      
I then discuss the case where the two matrices considered in Section $2$ are the real and imaginary
part of a complex matrix, and the form of the coupling to a transverse hermitean matrix. By means of a Bogoliubov
transformaton, the full string tension corrected
expression for the BMN energies is obtained.
I relate this result to the potential of an impurity in the presence of the single matrix backgound. 
Section $5$ is reserved for a brief discussion and conclusions.
  
In the Appendix, we describe how the local operators discussed in \cite{Donos:2005vm} are obtained from the general states introduced in
this article by a suitable projection, and discuss some of their properties. .

\noindent
\section{Free case and multi local fields}

\subsection{Two matrix Hamiltonian with a harmonic potential}

We consider first the free case, i.e., the Hamiltonian of two $N \times N$ hermitean matrices $M$ and $N$ in a harmonic potential. As is well 
known, the harmonic potential arises as a result of the curvature contribution in the leading Kaluza Klein compactification of the bosonic
sector of $\cal{N}=$ $4$ SYM on $S^3 \times R$. The exact physical identification of the matrices $M$ and $N$ will be discussed in the next section, 
although in general they will be linear combinations of two of the six adjoint scalar fields and their conjugate momenta. 

\noindent
The Hamiltonian is

\be
\hat{H}
\equiv
\frac{1}{2} Tr (P_M^2 ) + \frac{w^2}{2}  Tr (M^2) + \frac{1}{2} Tr (P_N^2 )+ \frac{w^2}{2}  Tr (N^2)
\ee

\noindent
with $P_M$ ($P_N$) canonical conjugate to $M$ ($N$, respectively). As suggested in \cite{Donos:2005vm}, we will
from now on use a coherent state representaton for the matrix $N$:\footnote{In this article, I will often
switch from creation-annihilation operators to their coherent state representation $B^{\dagger}\to B$,
$B\to\partial/{\partial B}$.}

\be \label{Hami}
\hat{H} = \frac{1}{2} Tr (P_M^2 ) + \frac{w^2}{2} Tr (M^2) + w  Tr (B {\partial \over \partial B})
\ee

\noindent
We are interested in the spectrum of gauge invariant states. One way to implement this invariance in the large $N$
limit, is to restrict the Hamiltonian to act on wave functionals of invariant single trace operators ("loops").

A complete set (in the large $N$ limit) of gauge invariant operators is given by:

\bea
\psi(k;0)  &=&  Tr(e^{ikM})   \nonumber\\
\psi(k;1)  &=&  Tr(B e^{ikM})\nonumber  \\
\psi(k_1,k_2;2)  &=&  Tr(B e^{ik_1 M} B e^{ik_2 M})\\
           & ... \nonumber \\
\psi(k_1,k_2,...,k_s;s)  &=&  Tr \big( \prod_{i=1}^s (B e^{ik_i M})\big), \quad s > 0. \nonumber 
\eea

Equivalently,

\bea\label{Densities}
\psi(x;0)&=& \int \frac{dk}{2\pi} e^{-ikx} \psi(k;0)  =  Tr(\delta(x-M))   \nonumber\\
\psi(x;1)&=& \int \frac{dk}{2\pi} e^{-ikx} \psi(k;1)  =  Tr(B\delta(x-M))    \\
\psi(x_1,x_1;2)&=& \int \int \frac{dk_1}{2\pi}\frac{dk_2}{2\pi} e^{-ik_1x_1} e^{-ik_2 x_2} \psi(k_1,k_2;2) \nonumber \\
&=&  Tr(B \delta(x_1-M) B \delta(x_2-M) )\nonumber \\
           & ... \nonumber \\
\psi(x_1,x_2,...,x_s;s)&=& \int ...\int \frac{dk_1}{2\pi}... \frac{dk_s}{2\pi}  
                        e^{-ik_1x_1} ... e^{-ik_sx_s}  \psi(k_1,k_2,...,k_s;s)\nonumber  \\
&=& Tr \big( \prod_{i=1}^s (B \delta(x_i -M) )\big), \quad s > 0. \nonumber 
\eea

\noindent
We will refer to these as "$s$ impurity states" and, to simplify notation, we will often denote them 
by $\psi(A;s)$, with $A$ an appropriate generic index. 

The restriction of the action of the hamiltonian (\ref{Hami}) on functionals of the invariant operators is
implemented by performing a change of variables \cite{Jevicki:1979mb} from the original matrix variables to
the invariant variables. Because of the reduction in the number of degrees of freeedom, 
the jacobian of this transformation has to be taken into account \cite{Jevicki:1979mb}. 
As loops with non-vanishing number of impurities have vanishing expectation values, this
Jacobian only depends on the zero impurity variables \cite{Avan:1995sp},\cite{Donos:2005vm}. 
Therefore, only the usual large $N$ single matrix background of the matrix $M$ is generated. 

\subsection{Background and zero impurity sector}

The zero impurity sector is nothing but the large $N$ Hamiltonian dynamics of a single hermitean matrix.  
This is completely described by the standard cubic collective field hamiltonian 
(in addition to the original derivation \cite{Jevicki:1979mb}
and its aplication to $c=1$ strings \cite{DasKA},\cite{Demeterfi:1991tz}, reference \cite{deMelloKoch:2002nq}
also has a general self-contained review of the method).    

Including only the terms required for a study of the background and fluctuations,
this hamiltonian takes the form:

\be\label{CollBef}
{-{1 \over 2}}
\int dx \partial_x {\partial \over \partial \psi(x,0)} \psi(x,0) \partial_x {\partial \over \partial \psi(x,0)}
+ \int dx \Big( {\pi^2 \over 6}\psi^3(x,0) +  \psi(x,0)({w^2 x^2 \over 2}- \mu) \Big)
\ee

\noindent
where the Lagrange multiplier $\mu$ enforces the contraint

\be \label{Const}
\int dx \psi(x,0) = N .
\ee

\noindent
To exhibit explicitly the $N$ dependence, we rescale

\bea \label{Rescaling}
x  &\to& \sqrt{N} x \nonumber\\
\psi(x,0) &\to& \sqrt{N} \psi(x,0) \nonumber\\
-i {\partial \over \partial \psi(x,0)}\equiv \Pi(x)  &\to& {1 \over N} \Pi(x) \\
\mu &\to& N \mu\ \nonumber
\eea

\noindent
and obtain

\bea \label{HEffZero}
H_{eff}^{0}&=& {{1 \over 2N^2}}
\int dx \partial_x \Pi(x) \psi(x,0) \partial_x \Pi(x) \\ 
&+& N^2 \Big(  \int dx {\pi^2 \over 6}\psi^3(x,0) +  \psi(x,0)({w^2 x^2 \over 2}- \mu) \Big), \nonumber
\eea

\noindent
giving the well known Wigner distribution background in the limit as $N \to \infty$:

\be\label{po}
\pi \psi(x,0) \equiv \pi \phi_0 = \sqrt{2\mu -w^2 x^2} = \sqrt {2 w - w^2 x^2}.
\ee

\noindent
The droplet picture emerges naturally in this formalism: if we let 
\cite{Avan:1991kq} $p_{\pm}\equiv \partial_x \Pi(x) / N^2 \pm \pi \psi(x,0)$, 
then (\ref{HEffZero}) has a very natural phase space 
representation as 

$$
H_{eff}^{0} = \frac{N^2}{2\pi} \int_{p_{-}}^{p_+} \int dx \big( \frac{p^2}{2} + \frac{x^2}{2} - \mu \big). 
$$ 

\noindent
As $N \to \infty$, the boundary of the droplet is given by $p_{\pm}^2 + x^2= 2\mu$.

For the small fluctuation spectrum, one shifts 

$$
\psi(x,0) = \phi_0 + {1\over \sqrt{\pi} N} {\partial_x \eta };
\qquad \partial_x \Pi(x) = - \sqrt{\pi} N P (x)
$$

\noindent
to find the quadratic operator

$$
H_{2}^{0}= {{1 \over 2}}
\int dx (\pi\phi_{0}) P^2(x) + {1 \over 2} \int dx (\pi\phi_0) ({\partial_x \eta })^2
$$

\noindent
By changing to the classical "time of flight" $\phi$

\be\label{phio}
{dx \over d\phi} = \pi \phi_0 ; \quad x(\phi)= - \sqrt{\frac{2}{w}} \cos(w\phi) ; \quad \pi \phi_0 =  \sqrt{2 w} \sin (w \phi);
\quad 0 \le \phi \le \frac{\pi}{w} ,
\ee

\noindent
one obtains the Hamiltonian of a free $1+1$ massless boson \cite{DasKA}:

\be\label{Quadratic}
H_{2}^{0}= {{1 \over 2}}
\int d\phi P^2(\phi) + {1 \over 2} \int d\phi ({\partial_\phi \eta })^2
\ee

Further imposition of Dirichelet boundary conditions at the classical turning points, for a consistent
time evolution of the constraint (\ref{Const}), yields the spectrum in the zero impurity sector 

\be \label{SpecZero}
                  \epsilon_j= w j  \quad ; \qquad     \phi_j = \sin(j w \phi),  \quad j=1,2 , ...
\ee

\noindent
The variable $\phi$ has a clear gravity interpretaton \cite{Donos:2005vm}, as the angular variable
in the plane of the droplet \cite{Lin:2004nb}. Finally, we note that 
the harmonic oscillator potential is special, in that the Wigner distribution background 
also safisfies the well known BIPZ \cite{BrezinSV} equation

\be \label{BIPZ}
\int dz {\phi_0(z) \over (x-z)}= w x .
\ee

\noindent
This result will turn out to be of importance in the following.

\subsection{Spectrum of states with impurities - coarse graining}

\noindent
In \cite{Donos:2005vm} it was shown that the form of the quadratic operator determining the many-impurity spectrum is

\be \label{HTwoMulti}
H_2^{s} = {-\frac{1}{2}} \sum_{A} \bar{\omega} (A,s) {\partial \over \partial \psi(A,s)}
+ {1\over 2} \int dx \sum_{A} \Omega(x,0:A,s)
          {\partial \ln J \over \partial \psi(x,0)} {\partial \over \partial \psi(A,s)}
\ee

\noindent
where, to leading order in $N$

\be \label{Jac}
\partial_x {\partial \ln J \over \partial \psi(x,0)} = \partial_x \int dy \Omega^{-1}(x,0;y,0) \omega(y,0) =
2 \int dy {\phi_0(y) \over (x-y)}
\ee

\noindent
In (\ref{HTwoMulti}) and (\ref{Jac}), $\omega (A,s)$ and $\Omega(x,0:A,s)$ have their usual meanings \cite{Jevicki:1979mb}:

\bea \label{LapK}
\omega (A,s)&=& Tr({\partial^2 \psi(A,s) \over \partial M^2}) \\
\Omega (x,0:A,s)&=& Tr({\partial\psi(0,x) \over \partial M}{\partial\psi(A,s) \over \partial M}).
\eea

\noindent
$\omega (A,s)$ splits the loop $\psi(A,s)$ and $\Omega (A,s:A',s')$ joins the two loops $\psi(A,s)$ and $\psi(A',s')$.
$\bar{\omega}(A,s)$ indicates that only splittings into zero impurity loops need be considered.

\noindent
We have:

\be
\Omega (k_0,0:k_1,...,k_s,s) = - k_0 (\sum_{i=1}^s k_i \psi(k_1,...,k_i+k_0,...,k_s;s))
\ee

\be
\bar{\omega}(\{k_i\}) = - 2 \sum_{i=1}^s \int_0^{k_i} dk' k' \psi(k_i-k';0)\psi(k_1,...,k_{i-1},k',k_{i+1},...,k_s;s)    
\ee

\noindent
Equivalently

\be
\Omega (z,0:x_1,...,x_s,s) = \sum_{i=1}^s \frac{\partial}{\partial z} \frac{\partial}{\partial x_i}
 ( \delta(z-x_i) \psi(x_1,...,x_i,...,x_s;s))
\ee

\bea
& &\bar{\omega}(\{x_i\})= - 2 \sum_{i=1}^s \int dz  \psi(z:0) \frac{\partial}{\partial x_i}
                            ( \frac {\psi(\{x_i\};s)}{x_i -z} ) \\
& &- 2  \sum_{i=1}^s  \int dz  \psi(z;0)\big[ \frac{\psi(\{x_i\};s)}{(x_i-z)^2}- 
                             \delta(z-x_i) \int dy_i \frac{\psi(x_1,...,y_i,...,x_s;s)}{(y_i-z)^2} \big] \nonumber
\eea

\noindent
Substituting these in (\ref{HTwoMulti}) and making use of (\ref{Jac}), we obtain for the quadratic many-impurity
operator

\bea\label{HTwoM}
&&H_2^{s} = \int dx_1 ...  \int dx_s \int dz  \\ 
&& \sum_{i=1}^s  \frac{\phi_0(z) \psi(\{x_i\};s)- \phi_0(x_i) \psi(x_1,...,z,...,x_s;s)}{(x_i-z)^2}
\frac{\partial}{\partial \psi(\{x_i\};s) } \nonumber
\eea

\noindent
The rescaling (\ref{Rescaling}) implies that 

$$
\psi(A,s)\frac{\partial}{\partial \psi(A,s)} \to \frac{1}{N^{\frac{s}{2}}} 
\psi(A,s)\frac{\partial}{\partial \psi(A,s)},
$$

\noindent
so that (\ref{HTwoMulti})(and (\ref{HTwoM})) is invariant, i.e., of order $1$ ($N^0$) in $N$. Writing it as

$$
\sum_A \sum_B \psi(A,s) K(A,B:s) \frac {\partial}{\partial \psi(B,s)},
$$

\noindent
we obtain

\bea
&&\sum_B \psi(A,s) K(A,B;s) \frac {\partial}{\partial \psi(B,s)}=
\sum_{i=1}^s \int dx_1 ... \int dx_s \psi(\{x_i\};s) \nonumber \\
&&\int dz  \frac{\phi_0(z)}{(x_i-z)^2}(
\frac{\partial}{\partial \psi(\{x_i\};s) }-
\frac{\partial}{\partial \psi(x_1,...,z,...,x_s;s) })
\eea

\noindent
The $s$ many-impurity kernel is then

\be\label{Kern}
K(\{x_i\},\{x'_i\};s)=
\sum_{i=1}^s \int dz  \frac{\phi_0(z)}{(x_i-z)^2}\Big( \big( \prod_{j\ne i} \delta(x_j-x'_j)\big)\big( \delta(x_i-x'_i) -  \delta(z-x'_i) \big)\Big)
\ee

\noindent
When acting on an eigenfunctional   

$$
          \Phi = \int ... \int dw_1 ... dw_s f(w_1,...,w_s)  \psi(w_1,...,w_s:s) ,
$$

\noindent
we obtain

\bea
&&\sum_{i=1}^s \int dz  \frac{\phi_0(z)}{(x_i-z)^2} (f(x_1,...,x_s)-(f(x_1,..,z,...,x_s) \\
&=& \sum_{i=1}^s \Big(  \big( -\frac{d}{dx_i} \int dz  \frac{\phi_0(z)}{x_i-z} \big) f(x_1,..,x_s)  
  + \frac{d}{dx_i} \int dz  \frac{\phi_0(z) f(x_1,..,z,...,x_s)}{x_i-z} \Big) \nonumber
\eea

\noindent
i.e., a sum of Marchesini-Onofri kernels \cite{MarchesiniYQ},\cite{Halpern:1981fj},\cite{Gross:1990md},\cite{Maldacena:2005hi},\cite{Donos:2005vm}.

\noindent
The eigenvalues and eigenfuntions of this operator follow from (\ref{BIPZ}) and the result

\be\label{Tcheb}
\int_{-\sqrt{\frac{2}{w}}}^{\sqrt{\frac{2}{w}}} \frac{dz}{\pi} \frac{\sin(n w \phi(z))}{x-z}= - \cos(n w \phi(x)),   
\ee

\noindent
which, for the harmonic potential, is related to a well known integral relationship between the 
two different kinds of Chebyshev polynomails. The eigenvalues and eigenfuntions are

$$
            \epsilon'_{n_i}= w (\sum_{i=1}^s n_i - s) \quad ; \quad  
\Psi_{n_1,...,n_s}^s (x_i)= \prod_{i=1}^s {\sin (n_i w \phi(x_i)) \over \sqrt{2 w} \sin(w \phi(x_i))} \quad ; \quad n_i=1,2,...
$$

\noindent
We recognize the Chebyshev polynomials of the second kind $U_{n_i-1}(w \cos(\phi))$. 
Adding the contribution from the $Tr (B {\partial / \partial B}) $ term of the Hamiltonian we obtain

\bea
   \epsilon_{n_i}=\sum_{i=1}^s n_i  \quad ; \quad 
 \Psi_{n_1,...,n_s}^s (x_i) &=&\prod_{i=1}^s {\sin (n_i w \phi(x_i)) \over \sqrt{2 w} \sin(w \phi(x_i))} \quad ; \quad n_i=1,2,...
\nonumber \\
                        &=& \prod_{i=1}^s {u_{n_i-1}(x_i)} \quad ; \quad n_i=1,2,..., 
\eea

\noindent
where to simplify notation I introduced the polynomials

\be\label{uu}
u_{n-1}(x) \equiv  {\sin (n w \phi(x)) \over \sqrt{2 w} \sin(w \phi(x))}=  {\sin (n w \phi(x)) \over \pi\phi_0(x)} =
 \frac{1}{\sqrt{2w}} U_{n-1}(-\sqrt{\frac{w}{2}} x)
\ee

We see a one to one correspondence with the spectrum of the states 
$\tr({A^{\dagger}}^{m_1} {B^{\dagger}}^{l_1}{A^{\dagger}}^{m_2}{B^{\dagger}}^{l_2}...)$, 
obtained by acting on the
original Fock space of the theory.

\subsection{Measure and Fock space}

\noindent
The states obtained in the previous subsection:

\be
            <\psi|\{n_i\},s> = \int_{-\sqrt{\frac{2}{w}}}^{\sqrt{\frac{2}{w}}} dx_1 ... \int_{-\sqrt{\frac{2}{w}}}^{\sqrt{\frac{2}{w}}} dx_s
                        \big{\{}  u_{n_1-1}(x_1)...u_{n_s-1}(x_s) \big{\}}_c  \psi (\{x_i\};s)
\ee 

\noindent
($\{...\}_c$ stands for cyclic symmetrization) form a complete orthonormal set with respect to the measure

\be\label{BigMes}
\int [d \psi ] \exp{\big[ -  \int \frac{dx_1}{\pi\phi_0}...\int \frac{dx_s}{\pi\phi_0} |\psi(\{x_i\};s)|^2 \big]}
\ee

\noindent
This is because the polynomials $u_n(x)$ form an orthonormal set with weight function $\pi\phi_0 = \sqrt {2 w - w^2 x^2} $.  

\noindent
In terms of the original matrix variables, these states take the form 

\be\label{Master}
  <\psi|\{n_i\}; s> = \tr (B u_{n_1-1}(M) B u_{n_2-1}(M)... B u_{n_s-1}(M) ).
\ee

\noindent
These states, with their well defined energies, 
provide the analogue of the usual states with well defined $U(1)$ charges built as traced products of complex monomials.

\noindent
They also have a more natural $\phi$ space representation, i.e with 

$$
            \psi(\phi_1,...,\phi_s;s) = \psi(x_1(\phi_1),...,x_s(\phi_s);s),
$$

\noindent
then

\be\label{Alphi}
 <\psi|\{n_i\}; s> =  \int_{0}^{\frac{\pi}{w}} d\phi_1 ... \int_{0}^{\frac{\pi}{w}} d\phi_s \sin(n_1 w \phi_1)... \sin(n_s w \phi_s) \psi(\{\phi_i\},s)
\ee

\noindent
and the measure (\ref{BigMes}) is trivial:

$$
\int [d \psi ] \exp{\big[ -  \int {d\phi_1}...\int {d\phi_s} |\psi(\{\phi_i\},s)|^2 \big]}
$$
 
\noindent
The orthonormality of the polynomials $u_n(x)$ is then seen to be a simple consequence of the orthogonality of the $\sin$ functions.  

\noindent
Equation (\ref{Alphi}) also exhibits a one to one correspondence between the eigenfunctions of multi-impurity states, 
and those of the zero-impurity sector, which are solutions of a $1+1$ dimensional massles Klein-Gordon equation. 
This is not entirely surprising, as it is known that the Marchesini-Onofri operator squares to the massless Klein-Gordon equation
\cite{Donos:2005vm} (and earlier references therein). Indeed, the expectation is that in a
full treatment where the second matrix is kept hermitean, all quadratic fluctuations will be determined by massless Klein-Gordon 
equations in the $\phi$ variable. 

\noindent
A very natural string Fock space suggests itself for the second quantized fields $\psi(\{x_i\};s)$ which can be constructed
along the lines of \cite{deMelloKoch:2002nq},\cite{deMelloKoch:2003pv}. This will not be pursued in this article.

\section {$g_{YM}$ interactions - Stringy hyperfine splitting and BMN spectrum}

\subsection{1/2 BPS variables}

We briefly review the precise identification of variables resulting from the dual description of 1/2 BPS states \cite{Corley:2001zk}. 
Starting with the leading Kaluza Klein compactification of the bosonic
sector of $\cal{N}=$ $4$ SYM on $S^3 \times R$, and choosing the plane defined by two scalars $X_1$ and $X_2$ grouped into a complex
matrix $Z=X_1+i X_2$, we introduce matrix valued creation and annhilation operators
\footnote{$w$ is inversely proportional to the $S^3$ radius, and it can be
scaled out of the action. It may be set to $1$ at the end of the calculation} 

$$
                                Z=\frac{1}{\sqrt{w}}  (A + B^{\dagger}) \qquad  \Pi =-i\frac{\sqrt{w}}{2}  ( A^{\dagger} - B).
$$

\noindent
$\Pi$ is the canonical conjugate to $Z$. 

Motion in this plane is characterized by the energy and the two dimensional angular momentum of the
free theory:

\be\label{Ei}
                           \hat{H_0}= w \big( \tr (A^{\dagger} A) +  \tr (B^{\dagger} B) \big) \quad 
                            \hat{J}=    \big( \tr (A^{\dagger} A) -  \tr (B^{\dagger} B) \big) .
\ee

\noindent
B carries a well defined quantum of charge $-1$, and 1/2 BPS states correspond to a restriction to the sector with no
$B$ excitations. Therefore $B$ is the impurity considered in \cite{Donos:2005vm}. $M$ is the hermitean matrix associated
with the $A$, $A^{\dagger}$ system \cite{Donos:2005vm}:

\be\label{Ide}
                             M \equiv \frac{1}{\sqrt{2w}}( A + A^{\dagger})  \quad P_ M = -i \sqrt{\frac{w}{2}}( A + A^{\dagger}) 
\ee

\noindent
We can also project out the B sector by taking a pp wave limit \cite{Berenstein:2002jq}, while retaining one the harmonic modes, denoted by $C$,
associated with one of the (complex) transverse directions \cite{deMelloKoch:2003pv}. The interaction hamiltonian splits into $F$ and $D$ terms. 
There are well known non-renormalization theorems that apply to the $D$ term contributions \cite{D'Hoker:2002aw},
so we will concentrate on the F term interaction hamiltonian
(\footnote{Throughout this article we consider all terms of the Hamiltonian to be normal ordered.}) 

\be\label{DT}
                       H_{int}= - \frac{g_{YM}^2}{w^2} \tr ([A^{\dagger},C^{\dagger}][A,C])
\ee

\noindent
The $C$ oscillator does not participate in the $Z$ plane $\hat{J}$ charge, but it contributes to the energy, so 
free states with $s$ $C$ impurities only can be classified as eigenstates of the operators (factors of $w$ in $\hat{J}$)

\bea
 \hat{J}= w  \tr (A^{\dagger} A) =  \frac{1}{2} Tr (P_M^2 ) + \frac{w^2}{2} Tr (M^2) &\quad&   j = \sum_{i=1}^s n_i - s \\
  \hat{H_0}= \frac{1}{2} Tr (P_M^2 ) + \frac{w^2}{2} Tr (M^2) +  w \tr (C^{\dagger} C) \big)  &\quad&  \epsilon_j = \sum_{i=1}^s n_i = j + s
  \nonumber
\eea

\noindent
where we have used the results of Section $2$. For instance, specializing to two impurities, all states with $j=j_1+j_2, j_1\ge0, j_2\ge0$
($j_i=n_i-1$) are degenerate.

\noindent
The analogue BMN \cite{Berenstein:2002jq} operators are now easily contructed

\bea\label{BMNdef}
          O^m_J &=& \frac{1}{\sqrt{J}} \sum_{j=0}^J  e^{\frac{2\pi i m}{J+1}j}<\psi| j,J-j;2>        \\
                &=&  \frac{1}{\sqrt{J}} \sum_{j=0}^J  e^{\frac{2\pi i m}{J+1}j} \tr (Cu_{j}(M)Cu_{J-j}(M)) \nonumber
\eea

\noindent
We now wish to calculate the first order perturbation theory correction to the free energy $J+2$ of
the BMN operator above. For this, we write (\ref{DT}) in terms of our hermitean variables (\ref{Ide}) after noticing that the rescaling
(\ref{Rescaling}) requires that we let $M \to \sqrt{N} M$. The interaction Hamiltonian then takes the form:

\bea\label{Hint}
 \hat{H}_{int}&=& - \frac{g_{YM}^2 N}{2 w} \tr( [M,C] [M,C^{\dagger}]) + i \frac{g_{YM}^2 }{2 w^2} \tr([C,C^{\dagger}][M,P_M] ) \\
  &-& \frac{g_{YM}^2 }{2 N w^3} \tr([P_M,C] [P_M,C^{\dagger}]) \equiv \hat{H_1}+\hat{H_2}+\hat{H_3}  \nonumber
\eea

\noindent
We obtain

\bea
&&\hat{H_1} \tr (Cu_{j}(M)Cu_{J-j}(M)) = \frac{g_{YM}^2 N}{w} \Big{\{} - 2 \tr (C M u_{j}(M) C M u_{J-j}(M)) \nonumber \\
&& + \tr (C M^2 u_{j}(M)Cu_{J-j}(M)) +  \tr (C u_{j}(M) C M^2 u_{J-j}(M))\Big{\}}.
\eea

\noindent
Using the identity

$$
U_{j+1}(x)-2 x U_{j}(x) + U_{j-1}(x)=0,
$$

\noindent
and keeping terms with $j_1+j_2=J$ only, as we are only interested in the first order correction to the energy,
and states with different $J$ are orthogonal as explained in Subsection ($2.4$), we obtain: 

\bea
&&\hat{H_1} <\psi| j,J-j;2>   = \frac{g_{YM}^2 N}{w^2} \big( 2 <\psi| j,J-j;2> \\
&&  - <\psi| j+1,J-j-1;2> - <\psi| j-1,J-j+1;2> \big).
\eea

\noindent
it follows from (\ref{BMNdef}) that the first order shift in the BMN loop energy coming from $\hat{H_1}$ is

\be\label{BMNEn}
\Delta \epsilon (O^m_J)= \frac{g_{YM}^2 N}{J^2} \big( \frac{2 \pi m}{w}\big)^2
\ee

\noindent
where as usual $\lambda'={g_{YM}^2 N}/{J^2}$ is kept finite as $J \to \infty$ 

\noindent
Typically, $\hat{H_2}$ and $\hat{H_3}$ will split the loop $<\psi| j,J-j;2>$ into states with $0+2$, $1+1$, $0+0+2$ and $0+1+1$ 
impurities. Because $0$ impurity states develop a background, $0+2$ and $0+0+2$ states can potentially correct the energy.
However, we are only interested in establishing if they generate a correction to the result (\ref{BMNEn}). 
From the $N$ dependence of $\hat{H_2}$ and $\hat{H_3}$ in (\ref{Hint}), it is clear that any other possible dependence on the
't Hooft coupling comes from the terms with $\tr 1 = N $. We have:

\bea
&&\hat{H_2} \tr (Cu_{j}(M)Cu_{J-j}(M)) = \frac{g_{YM}^2}{w^2} \Big{\{}
 2 N  \tr (C u_{j}(M) C u_{J-j}(M))  \nonumber \\
&& - \tr(u_{j})\tr (C^2 u_{J-j}(M)) - \tr(u_{J-j}) \tr (C^2 u_{j})\Big{\}} +{\rm interactions }.
\eea

\noindent
The terms proportional to the 't Hooft coupling are clearly not universral, as the only other finite parameter
 is $g_2=J^2/N$.

\noindent
For $\hat{H_3}$,we have

\bea
&&\hat{H_3} \tr (Cu_{j}(M)Cu_{J-j}(M)) = \frac{g_{YM}^2N}{w^3} \Big{\{} 2 \tr (Cu^{(1)}_{j}(M)Cu^{(1)}_{J-j}(M)) \nonumber \\
&& - (Cu^{(2})_{j}(M)Cu_{J-j}(M)) - (Cu_{j}(M)Cu^{(2)}_{J-j}(M))\Big{\}} + ...
\eea 

\noindent
where I have defined

\be
        u^{(1)}_j(x)= \frac{u_{j}-a_j^0}{x} \quad  u^{(2)}_j(x)= \frac{u_{j}-a_j^1 x -a_j^0}{x^2} \quad u_j(x)=  a_j^0 +  a_j^1 x +...+a_j^j x^j  
\ee

\noindent
Clearly, these polynomials can be written as linear combinations of Chebyshev polynomials of strictly lower order then $j$, so again
the above terms do not contribute, and the result (\ref{BMNEn}) is not corrected to this order. As is well well known,
this is the expected result. 

\noindent

It would be very interesting to develop the form of the string field theory emerging from this matrix
model along the lines of \cite{deMelloKoch:2003pv}. This is beyond the scope of this paper. 

\subsection{Real-Imaginary variables - BMN spectrum}

\noindent
We now consider the identification of the matrices $M$ and $N$ as the real and  imaginary part 
of a single complex matrix $Z$. Again, we will retain a typical term arising from the coupling to one of transverse
hermitean scalars denoted by $Q$:

\bea\label{RealTr}
\hat{H}_{int} &=& -  g_{YM}^2 N  \tr ([M,Q]^2) \\  
&=&  - \frac {g_{YM}^2N}{2w} \big( \tr ([M,C]^2 + 2 [M,C^{\dagger}][M,C] + [M,C^{\dagger}]^2 ) \nonumber
\eea

\noindent
We recognize the impurity number conserving term, up to a factor, as $\hat{H_1}$ in (\ref{Hint}). As explained 
in the previous subsection, its action on a BMN loop (\ref{BMNdef}) leads to a first order shift in energy:

\be\label{BMNEnt}
\Delta \epsilon (O^m_J)= 2 \frac{g_{YM}^2 N}{J^2} \big( \frac{2 \pi m}{w}\big)^2
\ee

\noindent
However, the form of the interaction in (\ref{RealTr}) suggests that it may be possible to 
eliminate the $C C$ and $C^{\dagger} C^{\dagger}$ terms by means of a Bogoliubov transformation

$$
     C_{ij} = \cosh (\phi_{ij}) \tilde{C}_{ij} - \sinh (\phi_{ij}) \tilde{C}^{\dagger}_{ij}  
$$ 

\noindent
This is indeed the case provided 

$$
 \tanh(2\phi_{ij}) = \frac{\frac {g_{YM}^2N}{w}(\lambda_i-\lambda_j)^2}{w + \frac{g_{YM}^2N}{w}(\lambda_i-\lambda_j)^2}
$$

\noindent
where the $\lambda_i$'s are the eigenvalues of the matrix $M$. Then

\bea\label{Htot}
\hat{H} &=& ... + \sum_{i,j=1}^N \sqrt{\big(w + \frac{g_{YM}^2N}{w}(\lambda_i-\lambda_j)^2\big)^2
 - \big(\frac {g_{YM}^2N}{w}(\lambda_i-\lambda_j)^2\big)^2} \quad\bar{C}^{\dagger}_{ij}\bar{C}_{ji}  \nonumber \\
  &=& ... + \sqrt{w^2 + 2 {g_{YM}^2N}(\lambda_i-\lambda_j)^2} 
\quad\bar{C}^{\dagger}_{ij}\bar{C}_{ji} 
\eea

\noindent
In the above, $\bar{C}=V^{\dagger}\tilde{C}V$, where $V$ is the unitary matrix that diagonalizes $M$ and $...$ denotes
normal ordering contributions.

\noindent
But

\bea
&&\sum_{i,j=1}^N \bar{C}^{\dagger}_{ij} \sqrt{w^2 + 2 {g_{YM}^2N}(\lambda_i-\lambda_j)^2} \bar{C}_{ij} =
w \tr (\tilde{C}^{\dagger} \tilde{C} ) \\
&& - \frac{ g_{YM}^2 N}{w} \tr ( [ M, \tilde{C}^{\dagger}] [M, \tilde{C}] ) 
- \frac{ (g_{YM}^2 N)^2}{2w^3} \tr ( [M,[M,\tilde{C}^{\dagger}]] [M,[M,\tilde{C}]]) + ... \nonumber
\eea

\noindent
The term linear in $g_{YM}^2N$ is the impurity conserving term in (\ref{RealTr}). 
As already discussed, the BMN 
loops diagonalize the action of this operator, so the above expansion simply completes the square root, and
on obtains 

$$
        \epsilon_m =  \sqrt    { w^2 + \frac{2 g_{YM}^2 N}{J^2} \big({2 \pi m}\big)^2 }
$$

\noindent
per impurity.


The above result can be understood as follows:  
the dynamics of the transverse hermitean scalar $Q$ in the background of the scalar $M$ is
determined by the potential

$$
                      V_Q =  \frac{w^2}{2} Tr (Q^2)  -  g_{YM}^2 N  \tr ([M,Q]^2)
$$ 

\noindent
The large N dynamics of $M$ is the large N dynamics of its eigenvalues described by the density of eigenvalues (\ref{Densities})
(\ref{Rescaling})

$$
        \frac{1}{N}         \sum_{i=1}^N \delta (x - \lambda_i) \equiv \psi (x,0) \to \phi_0 \quad {\rm as} \quad  N \to \infty.
$$

\noindent
Rewritting $V_Q$, as suggested in 
\cite{Berenstein:2005jq}, in the form 

$$
       V_Q =  \sum_{i,j=1}^N \frac{w^2}{2} \bar{Q}_{ij} \bar{Q}_{ji} +  g_{YM}^2 N  \bar{Q}_{ij}(\lambda_i-\lambda_j)^2 \bar{Q}_{ji},
$$

\noindent
shows that each of the cordinates $\bar{Q}_{ij}$  has a (background dependent) frequency $\bar{w}_{ij}$ given by 

\be\label{MatFreq}
     \bar{w}_{ij}^2 = w^2 +  2 g_{YM}^2 N (\lambda_i-\lambda_j)^2 ,                   
\ee

\noindent
in complete agreement with (\ref{Htot}).



This calculation is similar in spirit to that of 
\cite{Berenstein:2005jq} in the sense that there too the spectrum was calculated about a large N background,
but of commuting matrices. We also expand about a background, that of a single matrix, but there is no
need to assume commutativity.

\noindent

\section{Conclusion}

\noindent
We have sucessfully applied the idea that the large $N$ limit of a system of matrices with a harmonic
potential can be studied by treating one of them exactly and the others, in a creation/annihilation basis, 
in the background of the first, to two hermitean in the free case and then to BMN loops.   
The results of this aricle generalize in a straightforward way to the case of impurities of different types. 
This will be reported in a forthcoming publication \cite{ToHappen}.

\section{Acknowledgements}
I would like to thank Robert de Mello Koch, Antal Jevicki and Igor Klebanov for 
their comments on the manuscript. I would also like to thank the Princeton High Energy Theory Group for
affording me the possibility of spending part of my sabbatical in Princeton and for their hospitality.

\newpage
\section{Appendix: Free case - local states}

\noindent
In \cite{Donos:2005vm}, (super) gravity states $\psi_S(k,s)$ (or $\psi_S(x,s)$) were introduced:

\bea
\psi_S (k,1)&=& Tr(B e^{i k M} )= \psi (k,1) \nonumber \\
\psi_S (k,2)&=&\int_0^k dk_1 Tr(B e^{i k_1 M} B e^{i (k-k_1) M} ) = \int_0^k dk_1 \psi (k_1,k-k_1)\nonumber \\ 
\psi_S (k,3)&=&\int_0^k dk_2 \int_0^{k_2} dk_1 Tr(B e^{i k_1 M} B e^{i (k_2-k_1) M} B e^{i (k-k_2) M} ) \nonumber \\
&=&\int_0^k dk_2 \int_0^{k_2} dk_1 \psi(k_1, k_2-k_1,k-k_2;3) \nonumber \\
&  ...
\eea

\noindent
with  $\psi_S(x,s)$ obtained by Fourier transforms. The spectrum of these states was found in \cite{Donos:2005vm}, 
where it was also found to be linear. Here we explain how the same 
spectrum can be obtained from the more general states above, by a suitable projection.

\noindent
As an example, let us consider two impurities. One can easily establish that

\bea
\psi_S(k:2)&=& -i \int dx_1 \int dx_2 \psi(x_1,x_2:2)\int dk' e^{ik'x_1}e^{i(k-k')x_2} \nonumber \\
&=&  - 2 i \int dx_1 \int dx_2 \psi(x_1,x_2:2)  \frac{e^{ikx_1}}{x_1-x_2},
\eea

\noindent
and hence

$$
\psi_S(x:2)= - 2 i \int dy \frac{\psi(x,y:2)}{x-y}
$$

\noindent
We can now let the kernel (\ref{Kern}) act on states

$$
\int dx_1 g(x_1) \psi_S(x_1:2) = - i \int dx_1 \int dx_2 \big( \frac{g(x_1)}{x_1-x_2}- \frac{g(x_2)}{x_1-x_2} \big)  
$$ 

So instead of generic functions $f(x_1,x_2)$, the kernel acts on functions of the form  

$$
          f(x_1,x_2) =  \frac{g(x_1)}{x_1-x_2}- \frac{g(x_2)}{x_1-x_2}.
$$

\noindent
One obtains ($w=1$ in this discussion) 

\bea
\hat{K}[\frac{g(x_1)}{x_1-x_2}]&=& \frac{g(x_1)}{x_1-x_2} \Big( \int dz \frac{\phi_0(z)}{(x_1-z)^2}+  \int dz \frac{\phi_0(z)}{(x_2-z)^2} \Big) \\
&-& g(x_1) \int dz \frac{\phi_0(z)}{(x_2-z)^2}\frac{1}{x_1-z}
-  \int dz \frac{\phi_0(z)}{(x_1-z)^2}\frac{g(z)}{z-x_2} \nonumber
\eea

\noindent
By setting

$$
                           g(x) = \Phi_{n}(x)= \frac {\sin (n \phi(x))}{\sqrt{2} \sin(\phi(x))}, 
$$

\noindent
use of (\ref{BIPZ}), (\ref{Tcheb}) or their derivatives together with the use of partial fractions yields

$$
\hat{K}[\frac{g(x_1)}{x_1-x_2}] = -2 \frac{g(x_1)}{x_1-x_2}+\frac{\cos(n\phi(x_1))-\cos(n\phi(x_2))}{(x_2-x_1)^2} +  n \frac{ g(x_1)}{x_2-x_1}
$$

\noindent
Adding the contribution obtained by the cyclic permutation $1 \to 2$,$2 \to 1$, we obain:

$$
\hat{K}[\frac{g(x_1)-g(x_2)}{x_1-x_2} ]= (n-2)(\frac{g(x_1)-g(x_2)}{x_1-x_2})
$$

\noindent
This pattern generalizes. For instance, for three impurities, one takes 

$$
        f(x_1,x_2,x_3) =  \frac{g(x_3)}{(x_3-x_1)(x_3-x_2)}+ \frac{g(x_1)}{(x_1-x_2)(x_1-x_3)}+ \frac{g(x_2)}{(x_2-x_3)(x_2-x_1)}
$$

\noindent
and the eigenvalue is $n-3$. For $s$ impurities, and taking into account the  $Tr (B {\partial / \partial B}) $ term of the Hamiltonian we obtain

$$
        w_n= n  \quad ; \quad  \Phi_n^s (z)=  {\sin (nq(z)) \over \sqrt{2} \sin(q(z))} \quad ; \quad n=1,2,...
$$

\noindent
This confirms the results of \cite{Donos:2005vm}

\noindent
It is important to realize that these states are normalized with respect to measure different from the multi-local states discussed previously. For instance,
for two impurities, the measure is 

\be\label{SMes}
\int [d \psi_S] \exp{\big[ -  \int \frac{dx}{\pi\phi_0} |\psi_S(x;2)|^2 \big]} 
\ee

\noindent
but 

$$
 \int \frac{dx}{\pi\phi_0} |\psi_S(x;2)|^2 = 4 \int \frac{dx}{\pi\phi_0} \int dx_1 \int dx_2 \frac{1}{x-x_1} \frac{1}{x-x_2}\psi^*(x,x_1) \psi(x,x_2)
$$

\noindent
which is clearly distinct from (\ref{BigMes}), explaining the existence of another sequence of states.



\end{document}